\documentclass[12pt]{article}
\usepackage[dvips]{graphics}
\usepackage{amstex}
\usepackage{amssymb}

\begin{document}



\begin{center}
{\large \bf On  mixing angles and magnetic moment of heavy tau neutrino} 
\end{center}
\begin{center}
{\large  
S.N.Gninenko\footnote{E-mail address: Sergei.Gninenko\char 64 cern.ch}}
\end{center}

\begin{center}
{ Institute for Nuclear Research, Russian Academy of Sciences,\\ 
Moscow 117 312, Russia}  
\end{center}

\begin{abstract}
If the magnetic moment of unstable tau neutrinos with the mass of  $O$(MeV)
is in the region of $10^{-8}\mu_{B} \lesssim \mu_{\nu_{\tau}} \lesssim 10^{-6}\mu_{B}$, it is compatible with the 
present experimental and cosmological bounds. 

 It is pointed out here, that
if the tau neutrino has such a large magnetic moment and 
can oscillate into a neutrino of another flavour 
the results from $\nu e$ scattering experiment at LAMPF constrain the
tau neutrino 
mixing angles to  $sin^{2}2\theta_{e \tau} \lesssim 2\times 10^{-6} - 2\times  10^{-2}$ and
$sin^{2}2\theta_{\mu\tau} \lesssim 10^{-6} -  10^{-2}$ depending on the  magnetic moment value in the allowed region. 
\end{abstract}

\vspace{0.5cm}

In many extensions of the standard model the neutrino acquire a nonzero 
mass and a 
magnetic moment (for review see e.g. \cite{1}).\ Usually large magnetic moment imply large 
masses.\  In the simplest extension, for example, 
neutrino masses and magnetic moment are proportional \cite{2,3}.\
A nonzero mass of the neutrino is also required for Cosmology to solve 
the problem of dark matter in the Universe.\ 
 It is generally assumed that  massive tau neutrinos are natural candidates for the hot component of
the dark matter of the Universe, see e.g. ref. \cite{4}.\ But they cannot provide an explanation
for a cold dark matter, because of constraints from relic abundance.\
 However, as was proposed by Giudice \cite{5}, a stable (or quasi-stable in
cosmological times) tau neutrino 
with a mass in the range from $\simeq$1 MeV up to the present experimental 
upper limit of
 25 MeV and a magnetic moment $\mu_{\nu_\tau}$ as large as  
$\simeq 10^{-6}$ Bohr magneton ($\mu_{B} = e/2m_{e}$)
could be a possible candidate for a cold dark matter particle.\

Experimental limit from BEBC (CERN), 
$\mu_{\nu_{\tau}} < 5.4 \times 10^{-7}\mu_{B}$ \cite{6},
 and  revised  calculations of the evolution and freeze-out 
of the tau neutrino number density, ref. \cite{7}, definitely rule out  
Giudice's hypothesis.\ However, it was pointed out, that  
if the tau neutrino is an unstable particle with the lifetime exceeding one second, a band
of magnetic moment values 
\begin{equation}
10^{-8}\mu_{B} \lesssim \mu_{\nu_{\tau}} \lesssim 10^{-6}\mu_{B}
\end{equation}
remains compatible with experimental and cosmological bounds
\cite{7}.\

In this case, if the magnetic moment of $\nu_{\tau}$ exists,  
 massive tau neutrinos could manifest themselves in terrestrial 
experiments through the effect of 
$\nu_{\mu}(\nu_e)\rightarrow \nu_{\tau}$
neutrino oscillations \cite {8}, or, since 
magnetic moment value of $10^{-8}\mu_{B} \lesssim \mu_{\nu_{\tau}} \lesssim
10^{-6}\mu_{B}$ is  
large enough,
through tau neutrino  electromagnetic interactions, or through 
combined effect.\\

 In the recent paper, ref. \cite{9}, it was  shown that combined existence  of  $\nu_{\mu}\rightarrow \nu_{\tau}$
and/or $\nu_{e}\rightarrow \nu_{\tau}$
 oscillations and  a large nonzero magnetic moment of the tau neutrino would 
increase 
the total rate of events in $\nu_{\mu}(\nu_{e})$ neutrino- electron scattering experiments.\ The basic idea is rather simple.

Assuming 
 that a muon neutrino beam has a component of tau neutrinos due to 
$\nu_{\mu}\rightarrow \nu_{\tau}$ oscillations.\ 
In case of two-neutrino mixing
 neutrino states  evolve with a time $t$ as  

\begin{equation}
|\nu> (t) = a(t)|\nu_{\mu}>  + b(t)|\nu_{\tau}, \mu_{\nu_\tau}\neq 0>
\end{equation}    

where $|\nu_{\mu}>$ and $|\nu_{\tau}>$ denote  weak eigenstates 
of $\nu_{\mu}$ and $\nu_{\tau}$ neutrinos, and $a^{2}(t),~b^{2}(t)$ are the 
probabilities to find $\nu_{\mu}$ or $\nu_{\tau}$ in the beam at a given moment
 $t$, respectively.\ It is assumed that $a^{2}(0) =1$ at $t=0$.\  
The probability $b^{2}(t)$ 
depends on the parameters of $\nu_{\mu} - \nu_{\tau}$ oscillations as \cite{10}:   

\begin{equation}
b^{2}(t\simeq \frac{L}{c}) = P(\nu_{\mu}\rightarrow\nu_{\tau}) = 
sin^{2}2\theta_{\mu\tau} sin^{2}\frac{\Delta m^{2} L}{4E} 
\end{equation}
or
\begin{equation}
P(\nu_{\mu}\rightarrow \nu_{\tau}) \approx sin^{2}2\theta_{\mu\tau} sin^{2}\frac{1.27 \Delta m^{2}(eV^{2}) L(km)}{E(GeV)}
\end{equation}

where $sin^{2}2\theta_{\mu\tau}$ is the
 mixing angle, and 
$\Delta m^{2} = \bigl|m^{2}_{3} - m^{2}_{2}\bigr|$ 
 is the difference of squares of the mass eigenstates in $eV^{2}$, $E$ is the neutrino energy in GeV, and $L$ is
the mean distance between the neutrino source and the detector in km.\ 
In the above formulas it is also assumed that 
 magnetic field $B$ is weak enough not to affect the probability of
oscillations $P(\nu_{\mu}\rightarrow \nu_{\tau})$, i.e.  $\Delta m^{2}/2E \gg
\mu_{\nu_\tau}B$ \cite{11}.\

Then if magnetic moment of the
 $\nu_{\tau}$ exists,
it will contribute to a non-coherent part
of the $\nu_{\tau}e^{-}$ scattering cross section 
via the reaction that change the helicity of the tau neutrino (hence 
 right-handed neutrino states should exist).\ 
 This 
might result in observable deviations from purely  $\nu_{\mu}e^{-}$ 
electro-weak reaction which is well predicted by the SM.\
  Indeed, since the 
electromagnetic cross section is orders of 
magnitude larger than the weak cross section, even a small fraction of 
tau neutrinos with nonzero magnetic moment in the muon neutrino beam could lead
 to an observable effects in $\nu_{\mu}e^{-}$  scattering,
 while the magnetic moment 
of muon neutrino could be small enough to contribute effectively to 
$\nu_{\mu}e^{-}$ scattering.\
 The production rate of isolated electrons via 
$\nu_{\tau} e^{-}$  scattering in 
the detector depends also on the probability 
$P(\nu_{\mu}\rightarrow\nu_{\tau})$ to find a $\nu_{\tau}$ neutrino in the 
neutrino beam.\ This probability can be calculated from the neutrino 
survival and transition probabilities using Eqs.(3,4).\

This effect can be used to constrain mixing angles 
of the tau neutrino with neutrino of another flavours.\ It was shown, ref. \cite{9}, that using the results obtained from 
the experiment on study of the $\nu_{e}e^{-}$ elastic 
scattering at LAMPF, ref. \cite{12}, the following
  combined limits on the mixing angles and tau neutrino magnetic 
moment  for  $\Delta m^{2} \gtrsim 10~ eV^{2}$  can be derived

\begin{equation}
sin^{2}2\theta_{e\tau}\times \mu_{\nu_{\tau}}^{2} \leq 2.3 \times 10^{-18}\mu_{B}^{2} 
\end{equation}
\begin{equation}
sin^{2}2\theta_{\mu\tau}\times \mu_{\nu_{\tau}}^{2} \leq 1.1 \times 10^{-18}\mu_{B}^{2} 
\end{equation}

Note, that if  the tau neutrino mass is in the 
range from $\simeq 1$ to 25 MeV then
\begin{equation}
\Delta m^{2} = \bigl|m^{2}_{3} - m^{2}_{2}\bigr| \gg 10~ eV^{2}
\end{equation}
since the $m_{2}$ mass is experimentally measured to be smaller than 
160 KeV/$c^{2}$ \cite{13}.\
 
Assuming  the tau-neutrino magnetic moment to be in the range of 
$10^{-8}\mu_{B} \lesssim \mu_{\nu_{\tau}} \lesssim 10^{-6}\mu_{B}$
the following limits on the mixing angles  can obtained from Eqs.(5,6)

\begin{align}
sin^{2}2\theta_{e\tau} \lesssim 2\times 10^{-6} 
\hspace{1.0cm}      \mu_{\nu_{\tau}} \sim 10^{-6}\mu_{B}\\ 
sin^{2}2\theta_{e\tau} \lesssim 2\times 10^{-2}
\hspace{1.0cm}     \mu_{\nu_{\tau}} \sim 10^{-8}\mu_{B} 
\end{align}

\begin{align}
sin^{2}2\theta_{\mu\tau} \lesssim 10^{-6} 
\hspace{1.0cm} \mu_{\nu_{\tau}}\sim 10^{-6}\mu_{B}\\
sin^{2}2\theta_{\mu\tau} \lesssim 10^{-2}
\hspace{1.0cm} \mu_{\nu_{\tau}}\sim 10^{-8}\mu_{B}
\end{align}

The limits are valid for the mass region $\Delta m^{2} \gtrsim 10~ eV^{2}$ which is 
consistent with the mass region given by Eq.(7).\
They are obtained under assumption of combined existence of 
neutrino oscillations and a (large) magnetic moment of the tau neutrino.\ Nevertheless, it is interesting to compare them with
the present  experimental limits of 
$sin^{2}2\theta_{e\tau} < 0.15$ obtained from   
$\overline{\nu_{e}}\rightarrow \nu_{x}$ disappearance reactor experiments 
in  Bugey \cite{14} and in Krasnoyarsk \cite{15},
 or $ sin^{2}2\theta_{\mu\tau} < 0.01$
for $\nu_{\mu}\rightarrow \nu_{\tau}$, 
obtained from the exclusion plot of ref.\cite{16} for $\Delta m^{2} > 10~ eV^{2}$.\ For $\mu_{\nu_{\tau}}\sim 10^{-6}\mu_{B}$, the limit from Eq.(10) is also more stringent  than  the 
preliminary  experimental limits of 
$ sin^{2}2\theta_{\mu\tau} < 3.5 \times 10^{-3}$ from the CHORUS \cite{17}, or
 $ sin^{2}2\theta_{\mu\tau} < 3.7 \times 10^{-3}$ from the NOMAD \cite{18}
experiments at CERN obtained for the mass region $\Delta m^{2} > 1000~eV^{2}$.

On the other hand, if CHORUS or NOMAD will find $\nu_{\mu}\rightarrow \nu_{\tau}$
oscillations one would expect $ sin^{2}2\theta_{\mu\tau}\simeq 10^{-4}$ for 
the large $\Delta m^{2}$.\
 In this case, using Eqs.(1,6) one can obtain that a small window around $\mu_{\nu_{\tau}} \sim O(10^{-7})\mu_{B}$
might still be of interest for searching for tau neutrino magnetic moment.

\vspace{1.0cm}

{\large \bf Acknowledgements}\\

I thank N.V.~Krasnikov 
for useful discussions and G.Giudice for useful comment.


\newpage

\begin{thebibliography}{99}
\bibitem{1}
R.N.~Mohapatra and P.B.~Pal, in ``Massive Neutrinos in Physics and 
Astrophysics'', World Scientific, Singapore, 1991.
\bibitem{2}
B.W.~Lee and R.E.~Shrock, Phys. Rev. {\bf D16} (1977) 1444.
\bibitem{3}
W.~Marciano and A.I.~Sanda, Phys. Lett. {\bf B67} (1977) 303.
\bibitem{4}
S.~Sarkar, Preprint hep-ph/9602260.
\bibitem{5}
G.F.~Giudice, Phys. Lett. {\bf B251} (1990) 460;

G.F.~Giudice, Mod. Phys. Lett. {\bf A6} (1991) 851.
\bibitem{6}
A.M.~Cooper - Sarkar et al., Phys. Lett. {\bf B280} (1992) 153.
\bibitem{7}
D.~Grasso and E.W.~Kolb, Phys. Rev. {\bf D54} (1996) 1374;

L.H.~Kawano et al., Phys. Lett. {\bf B275} (1992) 487.
\bibitem{8}
B.~Pontecorvo, Zh. Eksp. Teor. Fiz. {\bf 33} (1957) 549; [JETP {\bf 6}(1958)429].
\bibitem{9}
S.N.~Gninenko, ''Limits on 
$\nu_{\mu}(\nu_{e}) \rightarrow \nu_{\tau}$  oscillations and  
tau neutrino magnetic moment from  neutrino-electron scattering'', Preprint hep-ph/9712220, to appear in Phys. Lett. B.
\bibitem{10}
See e.g. review: S.M.~Bilenky and B.~Pontecorvo, Phys. Rep.{\bf 41} (1978) 225.
\bibitem{11}
J.D.~Vergados , Nucl. Phys. Proc. Suppl. {\bf  22A}(1991) 21 .
\bibitem{12}
D.A.~Krakauer et al., Phys. Lett. {\bf B252} (1990) 177;

R.C.~Allen et al., Phys. Rev. {\bf D47} (1993) 11.
\bibitem{13}
K.~Assamagau et al., Phys. Lett. {\bf B335} (1994) 231.
\bibitem{14}
B.~Achkar et al., Nucl. Phys. {\bf 434} (1995) 503.
\bibitem{15}
G.S.~Vidyakin et al., JETP Lett. {\bf 59} (1994) 390.
\bibitem{16}
N.~Ushida et al., Phys. Rev. Lett. {\bf 57} (1986) 2897.
\bibitem{17}
CHORUS collaboration, CERN-PPE/97-149, submitted to 
Phys. Lett. B (1998);
\bibitem{18}
A.~Rubbia, ``Neutrino Physics at Accelerators'', Talk given at 
Lepton-Photon Symposium, Hamburg, 1997. 
\end{thebibliography}
\end{document}